\documentclass{PoS}
\let\ifpdf\relax
\usepackage{float}
\renewcommand{\refname}{\bf \large References} 

\renewcommand{\figurename}{\bf \footnotesize Fig.}

\title{\vspace{-40pt}\begin{flushright}{\footnotesize{LLNL-PROC-665000\vspace{24pt}}}\end{flushright}
Phase Structure Study of SU(2) Lattice Gauge Theory with 8 Flavors}

\ShortTitle{Phase Structure Study of SU(2) Lattice Gauge Theory with 8 Flavors}

\author{\speaker{Cynthia Y.-H. Huang}\\
        Institute of Physics, National Chiao-Tung University, Hsinchu 300, Taiwan\\
       E-mail: \email{cynthia1122.py01g@nctu.edu.tw}}
 \author{C.-J. David Lin\\
Institute of Physics, National Chiao-Tung University, Hsinchu 300, Taiwan\\
E-mail: \email{dlin@mail.nctu.edu.tw}}
\author{Kenji Ogawa\\ 
Institute of Physics, National Chiao-Tung University, Hsinchu 300, Taiwan\thanks{The affiliation was valid until November, 2013.}\\
\vspace{-12pt}
}
\author{Hiroshi Ohki\\
Kobayashi-Maskawa Institute for the Origin of Particles and the Universe (KMI), Nagoya University, Nagoya, Aichi 464-8602, Japan\\
E-mail: \email{ohki@kmi.nagoya-u.ac.jp}}
\author{Enrico Rinaldi\\
Lawrence Livermore National Laboratory, Livermore, California 94550, USA\\
E-mail: \email{rinaldi2@llnl.gov}}

\abstract{We present the investigation of the strong bare-coupling regime of SU(2) lattice gauge theory with 8 fermion flavors in the fundamental representation. The simulations are performed with unimproved staggered fermions and the plaquette gauge action. One bulk phase transition is observed through the measurement of the plaquette. The results of cold-start and hot-start simulations, as well as the hysteresis study, indicate the order of this transition can be weakly first order. Using the smeared Polyakov loops, and a method inspired by the constraint effective potential, we study the vacuum structure near the confining-deconfining phase transition. The Dirac operator eigenvalue spectrum is investigated, where further analysis is needed to clarify the properties of the chiral phase structure.}

\FullConference{The 32nd International Symposium on Lattice Field Theory\\
                 23-28 June, 2014\\
                 Columbia University New York, NY}

\begin{document}
\section{Introduction}
In SU($N_c$) gauge theory, there is a transition between the chirally-broken and the infrared-conformal systems at a particular number of fermion flavors, $N_f =\ N_f^{(c)}$. Theories with $N_f$ slightly below $N_f^{(c)}$ may develop nearly-conformal behavior. They can be used to construct models for dynamical electroweak symmetry breaking. For SU(2) gauge theories, the case of $N_f =\ 6$ has been shown to be chirally-broken~\cite{Appelquist:2013pqa}. A natural choice to approach $N_f^{(c)}$, therefore, is $N_f =\ $8. There have been some studies of SU(2) gauge theory with $N_f =\ 8$~\cite{Ohki:2010sr,Rantaharju}, but no concrete conclusion was made for the existence of an infrared fixed point. 

Phase structure study serves as the first step for understanding a theory. It provides vital information for performing lattice calculations. Furthermore, discoveries of new bulk phase structure~\cite{Cheng:2011ic} can appear in such process. Also, as conjectured in Ref.~\cite{deForcrand:2012vh}, properties of a gauge theory with many flavors in the strong bare-coupling regime may reveal its possible conformal behavior in the continuum limit.

In this article we present preliminary results of the work on the bulk phase structure for the case of $N_f =\ 8$. We use unimproved staggered fermions and the plaquette gauge action. Currently, we employ isotropic lattices and impose periodic boundary conditions in four directions for both gauge bosons and fermions. The smallest/largest bare fermion mass ($m_f$) and lattice volume analyzed so far are 0.005/0.015 and $6^4$/$24^4$ (preliminary), respectively. Gauge configurations are saved every ten HMC-trajectories.
\section{The phase structure study and results}
First, we measure plaquettes to identify the bulk phase transition. Then, smeared Polyakov loops are used to study the transition regarding confinement. In the final subsection, we present the eigenvalue density of the Dirac operator and investigate the chiral phase structure.
\subsection{The plaquette}
\begin{figure}[H]
\centering
\begin{minipage}[b]{0.464\linewidth}
\includegraphics[width=1.06\textwidth]{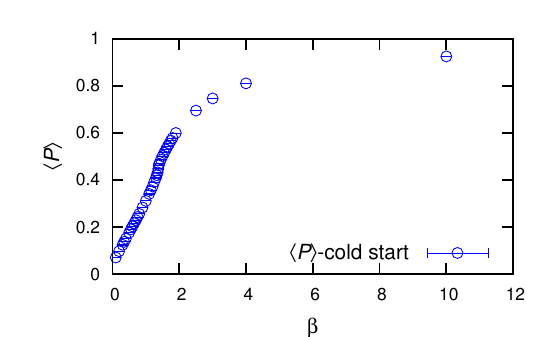}
\end{minipage}\\[-12pt]
\begin{minipage}[t]{\linewidth}
\vspace{-8pt}
\caption{\footnotesize The $\beta$-dependence of $\langle P\rangle$ for $m_f =\ $0.005 in the volume $8^4$.}
\vspace{-20pt}
\label{fig:1}
\end{minipage}
\end{figure}

\begin{figure}[H]
\vspace{-8pt}
\centering
 \begin{minipage}[b]{0.4905\linewidth}
\includegraphics[width=1.06\textwidth]{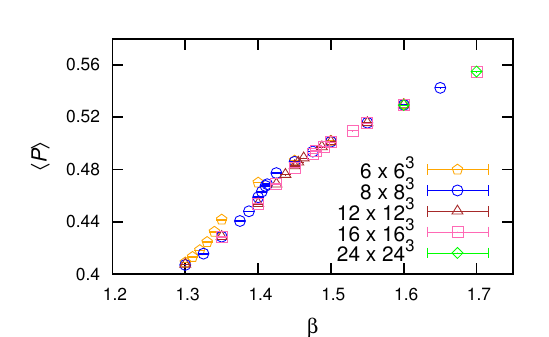}
\end{minipage}
\hspace{0.1cm}
\begin{minipage}[b]{0.4905\linewidth}
\includegraphics[width=1.06\textwidth]{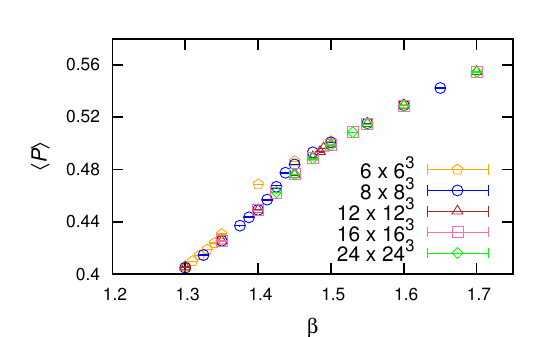}
\end{minipage}\\[-12pt]
 \begin{minipage}[t]{\linewidth}
 \vspace{-8pt}
\caption{\footnotesize The detail of $\langle P\rangle$ versus $\beta$ in the five volumes for $m_f =\ $0.010 (left) and $m_f =\ $0.015 (right).}   
\vspace{-5pt}
\label{fig:2}
\end{minipage}
\end{figure}

We define $P$ by averaging over all the plaquettes on $\mu\nu$-planes ($\mu <\nu$), normalized with the number of colors, $N_c$. As shown in Fig.~\ref{fig:1}, the discontinuity of $\langle P\rangle$ around $\beta\sim 1.4$ ($\beta=4\slash g^2$, where $g$ is the bare-coupling) indicates a phase transition. The detail near the transition is given in Fig.~\ref{fig:2}, where we observe no clearly-visible volume dependence when $N$ is larger than $12$ for $N^4$-lattice. This suggests that the phase transition is non-thermal. We perform cold-start and hot-start simulations, as well as the hysteresis study, to determine the order of this transition. As shown in
\begin{figure}[H]
\vspace{-15pt}
\centering
 \begin{minipage}[b]{0.4905\linewidth}
\includegraphics[width=1.06\textwidth]{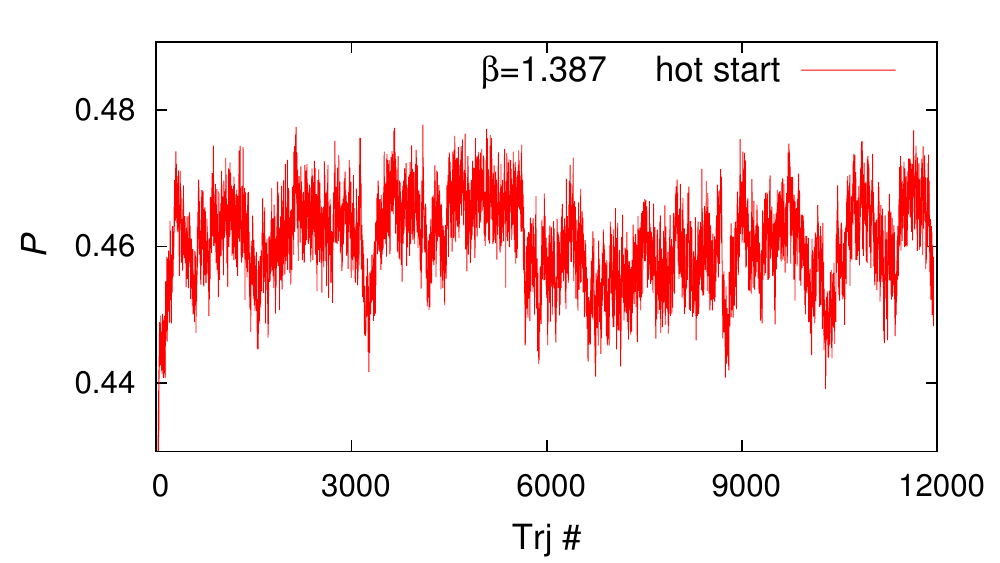}
\end{minipage}
\hspace{0.1cm}
\begin{minipage}[b]{0.4905\linewidth}
\includegraphics[width=1.06\textwidth]{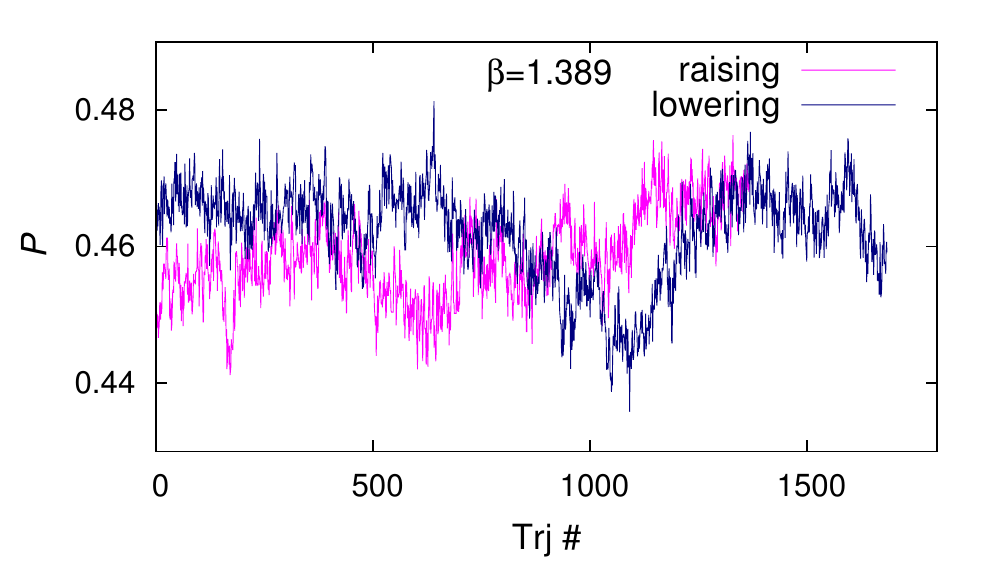}
 \end{minipage}\\[-12pt]
 \begin{minipage}[t]{\linewidth}
 \vspace{-8pt}
\caption{\footnotesize The HMC-histories of $P$ near the phase boundary for hot-start simulation at $\beta=$1.387 (left) and the hysteresis study at $\beta=$1.389 (right). Here $m_f =0.005$ and the volume is $8^4$.}   
\vspace{8pt}
\label{fig:ch}
\end{minipage}
\end{figure} 
\hspace*{-7.5mm}the left panel of Fig.~\ref{fig:ch}, the HMC-histories for hot-start simulation give indications of tunnelings near the phase boundary. In the hysteresis study, we increase/decrease $\beta$ by 0.001 every 100 trajectories. Between $1.375\leq\beta\leq1.4$ thirteen points are chosen, for which we continue the simulation after the adiabatic procedure. As shown in the right panel of Fig.~\ref{fig:ch}, the first 500 to 600 trajectories of the histories present evidence for coexistence of two states. This indicates that the bulk transition can be weakly first order, where further verifications are needed.
\subsection{Smeared Polyakov loops}
\begin{figure}[H]
\vspace{-8pt}
\centering
\begin{minipage}[b]{0.4908\linewidth}
\includegraphics[width=1.06\textwidth]{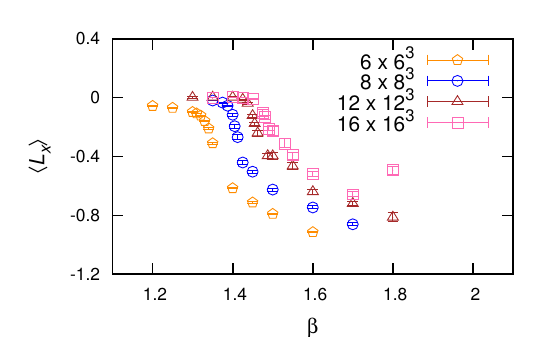}
 \end{minipage}
\hspace{0.1cm}
\begin{minipage}[b]{0.4908\linewidth}
\includegraphics[width=1.06\textwidth]{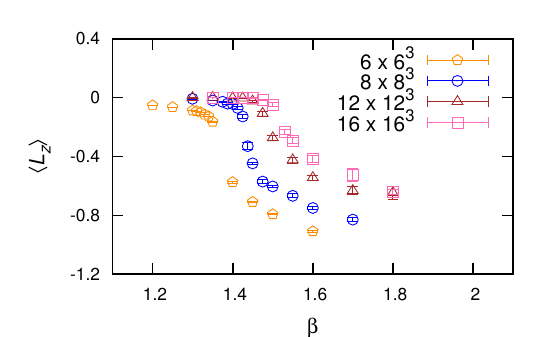}
 \end{minipage}\\[-12pt]
\begin{minipage}[t]{\linewidth}
\vspace{-8pt}
\caption{\footnotesize The detail of the $\beta$-dependence of $\langle L_x \rangle$ for $m_f$ $=\ $0.010 (left) and $\langle L_z \rangle$ for $m_f$ $=\ $0.015 (right) in the four volumes.}
\vspace{-5pt}
\label{fig:LcompV}
\end{minipage}
\end{figure}

\begin{figure}[H]
\vspace{-15pt}
\centering
\begin{minipage}[b]{0.461\linewidth}
\includegraphics[width=1.08\textwidth]{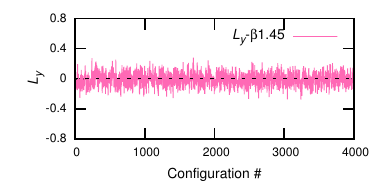}
  \end{minipage}
\hspace{0.1cm}
\begin{minipage}[b]{0.461\linewidth}
\includegraphics[width=1.08\textwidth]{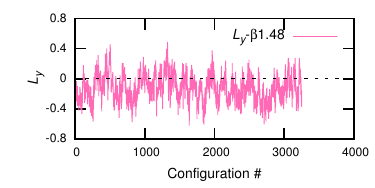}
 \end{minipage}\\[-10pt]
\begin{minipage}[b]{0.461\linewidth}
\includegraphics[width=1.08\textwidth]{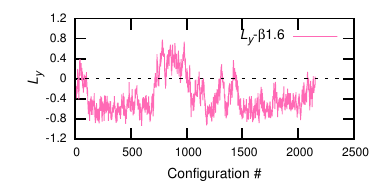}
 \end{minipage}
\hspace{0.1cm}
\begin{minipage}[b]{0.461\linewidth}
\includegraphics[width=1.08\textwidth]{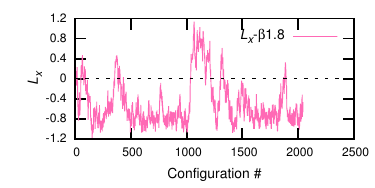}
 \end{minipage}\\[-12pt]
 \begin{minipage}[t]{\linewidth}
 \vspace{-9pt}
\caption{\footnotesize The HMC-histories of $L_y$ at $\beta=\ $1.45 (top left), $\beta=\ $1.48 (top right), $\beta=\ $1.6 (down left) and $L_x$ at $\beta=\ $1.8 (down right) for $m_f$ $=\ $0.010 and the volume $16^4$. Evidence of tunnelings is observed for the case of $\beta=1.48$.}   
\vspace{-5pt}
\label{fig:history4}
\end{minipage}
\end{figure}

\begin{figure}[H]
\vspace{-15pt}
\centering
\begin{minipage}[b]{0.47\linewidth}
\includegraphics[width=1.08\textwidth]{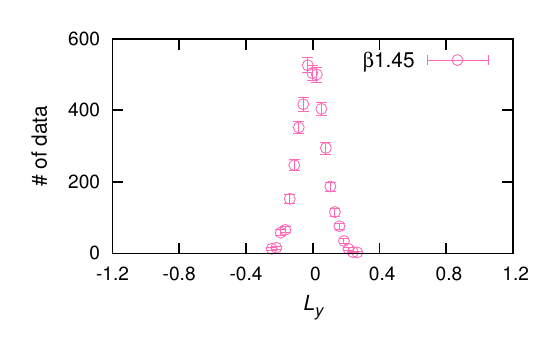}
\end{minipage}
\hspace{0.1cm}
\begin{minipage}[b]{0.47\linewidth}
\includegraphics[width=1.08\textwidth]{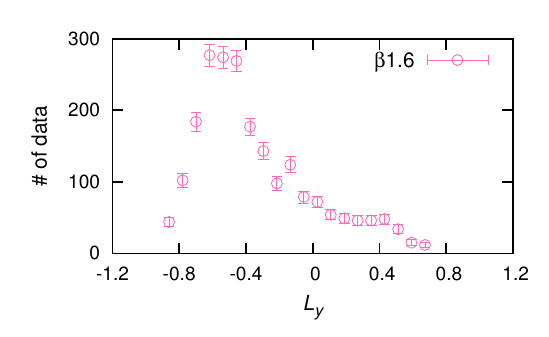}
\end{minipage}\\[-12pt]
\begin{minipage}[t]{\linewidth}
\vspace{-15pt}
\caption{\footnotesize The distributions of $L_y$ at $\beta=\ $1.45 (left) and $\beta=\ $1.6 (right) for $m_f$ $=\ $0.010 and the volume $16^4$. We have performed the bootstrap method to estimate the error of the distributions.}
\label{fig:histogramL}
\vspace{8pt}
\end{minipage}
\end{figure}

We measure Polyakov loops, $L_i$ ($i=\ x, y, z, t$), with APE smearing. To avoid finite-volume effects, we choose the smearing step ($N- $2) for $N^4$-lattice. Since there is no preferred direction for the isotropic lattices, the results of $L_i$ shown in this work are the exemplary cases which best present the phase transition concerning confinement. Figure~\ref{fig:LcompV} shows $\langle L_i \rangle$ as a function of $\beta$ near the phase transition point. The discontinuity of $\langle L_i \rangle$ separates the confining phase, where $\langle L_i \rangle$ $\simeq$ 0, and the deconfining regime, in which $\langle L_i \rangle <\ 0$. Figure~\ref{fig:history4} presents the HMC-histories of $L_i$. The fluctuation of $L_y$ is small in the symmetric phase while there are indications of tunnelings around the phase transition point. After the center symmetry is broken, $L_y$ and $L_x$ tunnel between the two vacua at $\beta=\ $1.6 and 1.8, respectively. As shown in Fig.~\ref{fig:histogramL}, the distribution of $L_y$ is symmetric in the confining phase. The skewed distribution in the deconfining regime at $\beta=1.6$ tends to smooth out the phase boundary. This can lead to difficulties for identifying the order of the transition. Pursuing a better way to determine it, we turn to a method inspired by the constraint effective potential.
\subsection{The constraint effective potential}
Let us first consider a single scalar field theory. The constraint effective potential (CEP), $V_{c}(\phi_{c})$, is defined through~\cite{Fukuda:1974ey} 
\begin{equation}
\mathbf{e}^{-\Omega V_{c}(\phi_{c})}= \mathit{const.} \int D\phi\, \mathbf{e}^{-\mathit{S}[\phi]} [\delta (\phi_{0}-\phi_{c})],
\label{eq:result}
\end{equation}
where $\phi_0$ is the zero four-momentum mode of the field, and $\Omega\equiv \int d^4 x$. It can be shown that~\cite{Fukuda:1974ey}, in the infinite volume limit ($\Omega\rightarrow\infty$) the CEP is identical to the effective potential. In numerical simulations, one can obtain the CEP by investigating the histogram for the zero modes of the scalar fields.

\begin{figure}[H]
\vspace{-5pt}
\centering
\includegraphics[width=0.66\textwidth]{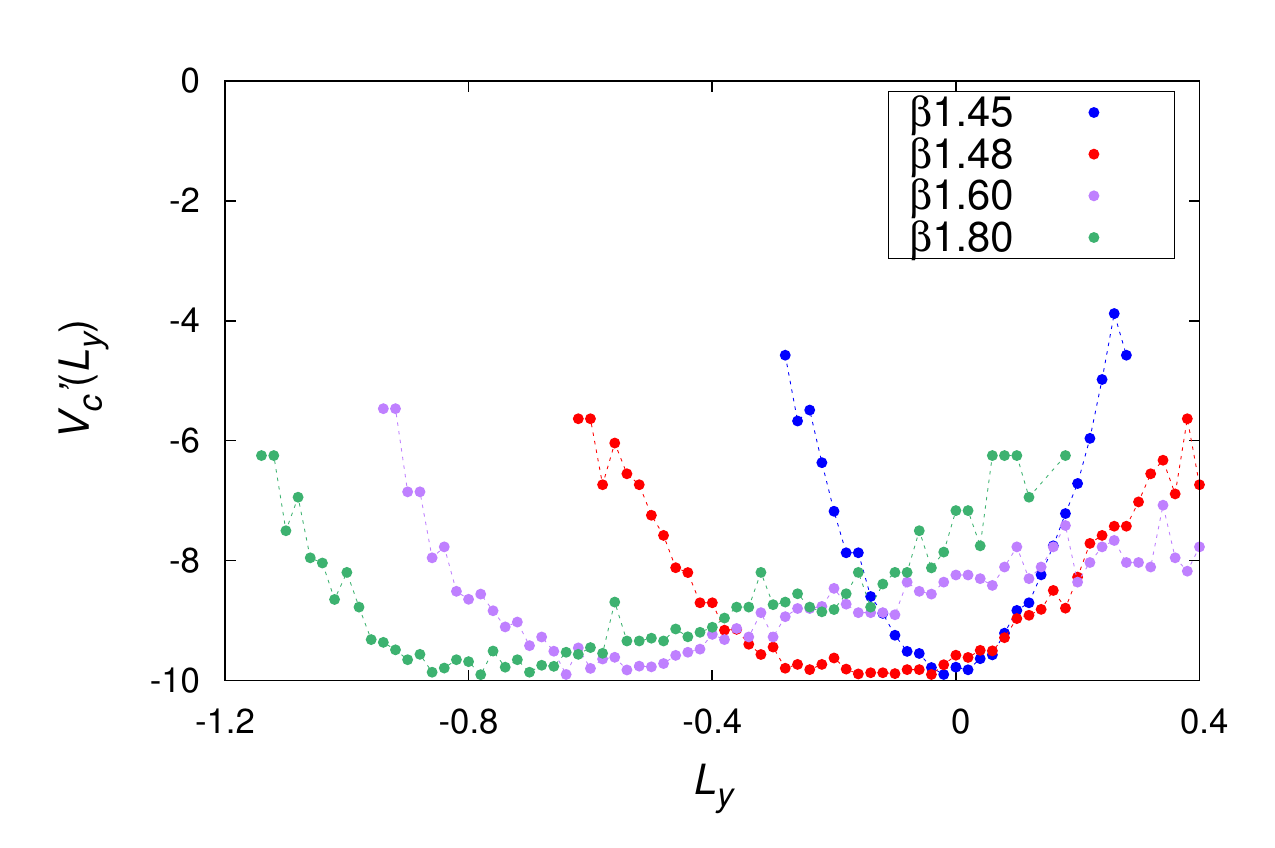}
\vspace{-12pt}
\caption{\footnotesize The CEP-inspired potential for $L_y$ at $m_f$ $=\ $0.010 and volume $16^4$ for $\beta=\ $(1.45, 1.48, 1.6, 1.8). The bin size $\Delta L$ is 0.02 in the histogram procedure. We have shifted $V_{c}'$($L_y$) along the $V_{c}'$-axis to make the minimum of the potentials to be the same.}
\vspace{8pt}
\label{fig:cons}
\end{figure}

Inspired by the CEP discussed above, we study the vacuum structure around the phase transition point by making the histogram for $L_i$ (We have fixed the bin size, $\Delta L= 0.02$, for all datasets):
\vspace{-8pt}
\begin{equation} 
V_{c}'(\mathit{L_i})=-\ln({\bf N})+C,
\end{equation}
with $V_{c}'(\mathit{L_i})$ the CEP-inspired potential, $\bf N$ the number of data points between $L_i$ and $L_i +\Delta L$, and $C$ depending on both the total number of configurations in a particular dataset and $\Delta L$. Near the phase boundary a double-well potential indicates that the transition is first order. On the other hand, a gradually-broadened one, in principle, corresponds to a second order phase transition. Figure~\ref{fig:cons} presents $V_{c}'(\mathit{L_y})$ at various choices of $\beta$. The value of $L_y$ which corresponds to the minimum of the potential moves away from zero as $\beta$ increases. At $\beta=\ 1.48$, the shape of the potential is different from the others: It is neither perfectly symmetric nor skewed-and-asymmetric. We infer that, at this point the simulation is very close to the phase boundary. Combining this observation with the study of the HMC-histories, where tunnelings happened, we conclude that the transition may be weakly first order.

\subsection{The Dirac operator eigenvalue spectrum}
\begin{figure}[H]
\vspace{5pt} 
\centering
\begin{minipage}[b]{0.465\linewidth}
\includegraphics[width=1.08\textwidth]{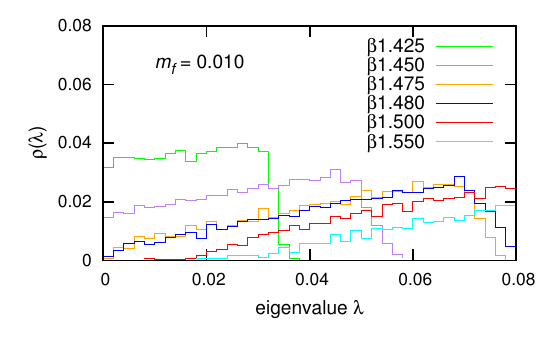}
\end{minipage}
\hspace{0.1cm}
\begin{minipage}[b]{0.465\linewidth}
\includegraphics[width=1.08\textwidth]{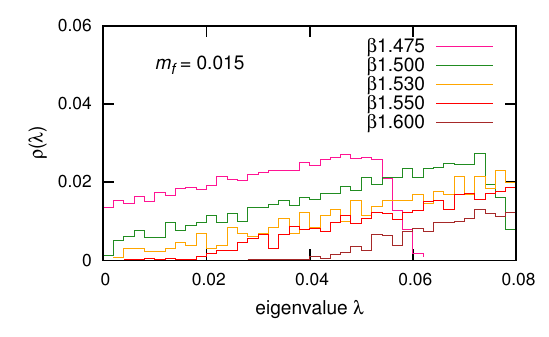}
\end{minipage}\\[-12pt]
\begin{minipage}[t]{\linewidth}
\vspace{-12pt}
\caption{\footnotesize The $\beta$-dependence of $\rho(\lambda)$ in volume $16^4$ for $m_f$ $=\ $0.010 (left) and 0.015 (right).}  
\vspace{-6pt} 
\label{fig:eigen1}
\end{minipage}
\end{figure}

\begin{figure}[H]
\vspace{-10pt}
\centering
\begin{minipage}[b]{0.465\linewidth}
\includegraphics[width=1.08\textwidth]{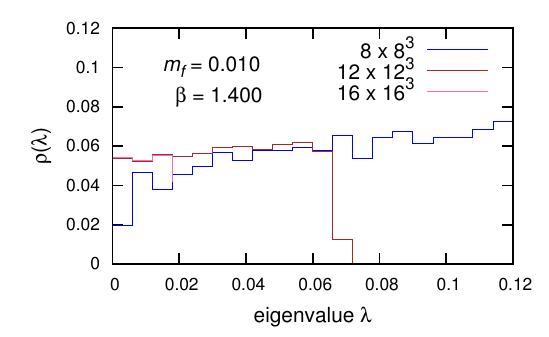}
\end{minipage}
\hspace{0.1cm}
\begin{minipage}[b]{0.465\linewidth}
\includegraphics[width=1.08\textwidth]{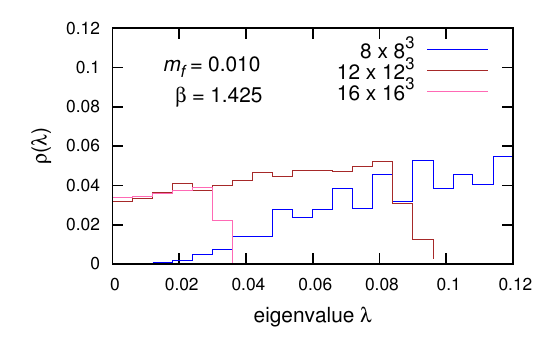}
\end{minipage}\\[-12pt]
\begin{minipage}[t]{\linewidth}
\vspace{-8pt}
\caption{\footnotesize The volume dependence of $\rho(\lambda)$ at $m_f$ $=\ $0.010 for $\beta=1.4$ (left) and $1.425$ (right).}
\vspace{-6pt} 
\label{fig:eigen2}
\end{minipage}
\end{figure}

\begin{figure}[H]
\vspace{-10pt}
\centering
\begin{minipage}[b]{0.465\linewidth}
\includegraphics[width=1.08\textwidth]{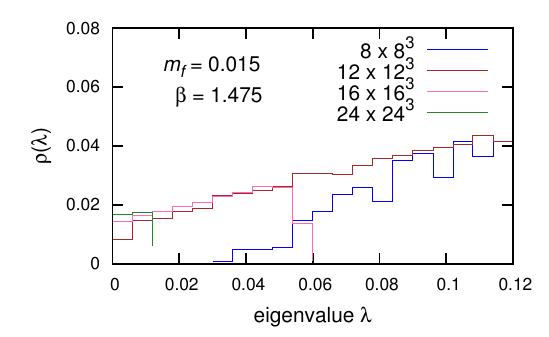}
\end{minipage}
\hspace{0.1cm}
\begin{minipage}[b]{0.465\linewidth}
\includegraphics[width=1.08\textwidth]{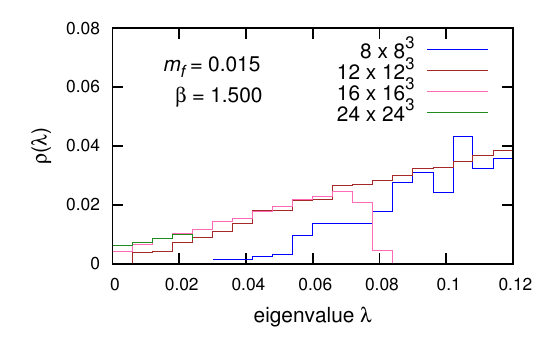}
\end{minipage}\\[-12pt]
\begin{minipage}[t]{\linewidth}
\vspace{-8pt}
\caption{\footnotesize The volume dependence of $\rho(\lambda)$ at $m_f$ $=\ $0.015 for $\beta=1.475$ (left) and $1.5$ (right).}
\label{fig:eigen5}
\end{minipage}
\end{figure}

We study the Dirac operator eigenvalue spectrum as the first step to computing the chiral condensate~\cite{Banks:1979yr, Marinari:1981nu, Leutwyler:1992yt}
\begin{equation}
\langle\bar{\psi}\psi\rangle=\ \lim_{m\to 0}\lim_{V\to\infty}[ -\pi \rho(\lambda=0)],
\end{equation}
where $\rho(\lambda)$ is the eigenvalue density in finite volumes. What follow are the preliminary results for both $\beta$-dependence and volume-dependence of $\rho(\lambda)$. Figure~\ref{fig:eigen1} shows examples for non-zero $\rho$(0) (gap) at the smaller $\beta$, which corresponds to non-vanishing condensate in the finite volume. As illustrated in Figs.~\ref{fig:eigen2} and~\ref{fig:eigen5}, the volume dependence of the gap increases as $\beta$ becomes larger, while it decreases when enlarging the volume for a fixed $\beta$.

There are two possible explanations for the onset of zero/non-zero gap in Fig.~\ref{fig:eigen5}. The volume dependence we observed can be consistent with a bulk phase transition. However, presently we are unable to exclude the possibility that the chiral-symmetry restoration is due to finite volume effects. Now, if the former is the case, there are two transitions regarding confinement and chiral symmetry breaking, and their phase boundaries are nearly-overlapped. In order to clarify the nature of the chiral phase transition, we need further investigations.


\section{Conclusion and outlook}
We have probed the phase structure of the SU(2) lattice gauge theory with $N_f =\ $8, which is a potential candidate for walking technicolor~\cite{Holdom, Yamawaki, Appelquist}. One bulk transition, which can be weakly first order, is identified through measuring the plaquette. The weakly first order confining-deconfining phase transition is discerned by using smeared Polyakov loops. We also observed a chiral phase transition in our preliminary investigation for the Dirac operator eigenvalue spectrum. However, we need more detailed analysis to determined the nature of this transition.

One can analyze the susceptibility for the plaquette and its volume dependence to confirm the order of the bulk transition. To extract $\langle\bar{\psi}\psi\rangle$, one needs to compute the chiral condensate in finite volumes at non-vanishing fermion masses and compare the results with the predictions from chiral Random Matrix Theory. These will be carefully studied and reported in our future publications. 

\acknowledgments{{\small We are indebted to Hideo Matsufuru for his help in developing the HMC simulation code. We thank A. Hasenfratz, I. Kanamori and A. Nagy for very helpful discussions. A. Nagy has kindly helped in the coding for computing the constraint effective potential. We are grateful to I. Kanamori for reading through the manuscript carefully and offering many useful comments. Simulations were done on the machines at National Center for High-Performance Computing in Taiwan and the computing system $\varphi$ at Kobayashi-Maskawa Institute, Nagoya University in Japan. C.Y.-H.H. and C.-J.D.L. are supported by Taiwanese Ministry of Science and Technology via grant number 102-2112-M-009-002-MY3. We acknowledge help in travel fund from National Centre for Theoretical Sciences, and C.Y.-H.H. is thankful for the travel fund from the Ministry of Education via grant number 103W981. H.O. is supported by the JSPS Grant-in-Aid for Scientific Research (S) No.22224003, and for Young Scientists (B) No.25800139. E.R. acknowledges the support of the DOE under contract DE-AC52-07NA27344 (LLNL). }}

\end{document}